# JOURNALS AS CONSTITUENTS OF SCIENTIFIC DISCOURSE: ECONOMIC HETERODOXY

Wilfred Dolfsma [1] & Loet Leydesdorff [2]

[1] University of Groningen,
School of Economics and Business,
PO Box 800,
9700 AV Groningen,
The Netherlands.
t. +31-50-3632789
f. +31-50-3637110
e. w.a.dolfsma@rug.nl

[2] Amsterdam School of Communications Research (ASCoR),
University of Amsterdam,
Kloveniersburgwal 48,
1012 CX Amsterdam,
The Netherlands.
t. +31-20-5256598
f. +31 20-5253681
e. loet@leydesdorff.net
http://www.leydesdorff.net/

JOURNALS AS CONSTITUENTS OF SCIENTIFIC DISCOURSE: ECONOMIC HETERODOXY

**Structured abstract**

**Purpose**: to provide a view and analysis of the immediate field of journals which surround a number of key heterodox economics journals.
**Design/methodology/approach**: Using citation data from the *Science* and *Social Science Citation Index*, the individual and collective networks of a number of journals in this field are analyzed.
**Findings**: The size and shape of the citation networks of journals can differ substantially, even if in a broadly similar category. Heterodox economics cannot (yet) be considered as an integrated specialty: authors in several journals in heterodox economics cite more from mainstream economics than from other heterodox journals. There are also strong links with other disciplinary fields such as geography, development studies, women studies, etc.
**Research limitations/implications**: The analysis is limited by its reliance on citation data in the *Science* and *Social Science Citation Indexes* provided by the Thomson-Scientific.
**Practical implications**: The analysis shows not only where journals draw their strengths from, but also how knowledge between journals and neighbouring sub-fields is diffused. This can be important for editors, authors, and others.
**Originality/value**: A network analysis focusing not just on a single journal as a focal point, but combining several journals in a single analysis enables us to visualize structural properties of the field of heterodox economics which otherwise remain latent. This study provides a structural approach to citation analysis as a tool for the study of scientific specialties.

## 1. Introduction

Increasingly journals have become the forum for scientific research to be presented and discussed so that contributions can be evaluated. This has long been the case for the natural sciences, and is increasingly so for the social sciences (Price, 1970). Books and edited volumes are losing much of the terrain they used to have in the not so distant past. Increasingly, decisions taken by university, faculty, and department boards regarding the hiring and promotion of academic staff, articles published in journals are used as a proxy for the quality of research. In some countries, research funds are allocated to universities on this basis. Some journals are more equal than others, however, and so the focus has moved to journals included in the *Social Science Citation Index (SSCI)* and the *Science Citation Index (SCI)*. Aggregated data is collected in the *Journal Citations Reports (JCR)* about citations from articles in these journals to articles in other journals also included. This data can be used to construct measures to characterize individual journals.

In this article we use this data not to rank journals in terms of impact factors and the like, but rather to understand their environment and their position in it. For this purpose, one can analyze the patterns of citations *from* as well *to* a given journal, that is, both the *citing* and the *cited* patterns, respectively. 'Citing' indicates how authors who publish in a journal relate to their larger academic environment; what knowledge they take from where. 'Cited' indicates how the relevant academic environment relates to the journal under study. For our purposes, we focus on citations *to* a journal, that is, the citation impact environment of the journal.

## 2. Journals, Citations, Reputations

A scientific community can, following Bourdieu (1992) be likened to a 'field' where relations and behaviour are institutionalized (Whitley, 1984). Academic activity is increasingly institutionalized in journals (Price, 1965), replacing associations and their conferences as the primary form of institutionalization. Journals are increasingly considered as an important way in which associations can create a platform for discussions in a field. If sufficiently broadly recognized, so that subscriptions in general and to libraries in particular accumulate, journals can be an important source of revenue for associations.

Another reason for journals to become the central locus of academic recognition for research is the seemingly more objective way in which publications for journals are selected (Snodgrass, 2006). Most journals of esteem nowadays use a system called double-blind peer-review whereby editors select reviewers for a submitted paper who evaluate an anonymous copy of the paper. These reports are used by editors to make their decisions. Decisions plus unsigned versions of the reports are then returned to the authors, who may be asked to revise and resubmit. One indicator that this system is more objective is the fact that more junior scholars and women are published in journals that use such a system as compared to a system where an editorial board decides (Blank, 1991; Goldberg, 1968). Other things equal, authors at prestigious academic institutions will have their papers accepted for publication more readily than near-top universities and non-academic institutions (Blank 1991).

Impact factors for journals are increasingly used to discriminate between publications produced by scholars, in addition to distinguishing between journal articles, on the one hand, and books or book chapters, on the other. Some departments, for example, recognize only articles published in journals that are ISI listed and thus have an impact factor. Some departments use a minimum IF for a journal to be recognized. Since the impact factor is based on a highly skewed distribution and limited to a two-year citation window, these evaluation practices raise validity issues (Leydesdorff, 2008).

As the need to evaluate scientific efforts and output has grown (Gibbons *et al*., 1994), a number of measurements have been proposed for journals. One is the total number of citations to a journal over the years. In a groundbreaking article, Garfield (1972) proposed the impact factor of a journal as a measure of its importance. This proposal has, needless to say, provoked a lot of discussion (Moed, 2005; Bensman, 2007; Pendlebury, 2007; Rossner *et al*., 2008a and b). Taking the total number of citations as a measure of 'prestige' disfavours newly established journals. Another measure, also based on citation data, is the *h-index* (Hirsch, 2005). This measure can be applied at the level of individual scientists too, and is best explained in that context. A scholar with an index of *h* has published *h* papers each of which has been cited by others at least *h* times. The *h*-index measures the set of the scientist's most cited papers and the number of citations they have received in other people's publications. Obviously this indicator also favours accumulation over time.

Citation patterns are now recognized as field specific (Garfield 1980). Journals in the natural sciences, e.g., tend to have far higher impact factors than journals in the social sciences for a number of reasons. Among the social sciences there are important differences as well (Price 1970). Even citation patterns between sub-fields within a single field may differ substantially (Leydesdorff 2008), sometimes following a business cycle of what topics are in demand. Some journals may also be more focused on the cutting edge of knowledge development in their field, or focus rather on reviewing the state of the art. Authors may also seek to relate to other fields by publishing in journals that bridge between fields (Goldstone & Leydesdorff, 2006). Comparing IFs across academic fields should thus be done with great care.

Most measurements aggregate citation data provided by Thomson / ISI in order to characterize a particular aspect of the position of a journal (or an author) in its relevant environment. In this study, we want to take a more comprehensive view of a journal's position in its field, using citation data as well. We use approaches from social network analysis and scientometrics for this purpose. In addition, betweenness centrality will be used as an indicator of a journal's interdisciplinarity (Leydesdorff 2007b).

## 3. Data, Measures, and Methods

We have combined the data from the *Social Science Citation Index (SSCI)* and the *Science Citation Index (SCI)*, provided by Thomson Scientific, formerly known as Thomson ISI, in order to analyze the field of heterodox economics. The journals chosen for the analysis could well be cited by or cite journals from the social sciences and the sciences, and so data from CD-Rom versions of the *Journal Citations Reports (JCR)* of both the *Science Citation Index* as well as the *Social Sciences Citation Index* were collected. The ISI staff aggregates data among journals in these *Journal Citations Reports*. For the year 2005 the two databases covered 5,968 and 1,712 journals, respectively. Since 301 journals are covered by both databases, a citation matrix can be constructed for (5,968 + 1,712 – 301) = 7,379 journals. This data can be analysed to provide both a snapshot and an understanding of the development of the relevant environments of a given journal.

The *JCRs c*ontain three main indicators of journals: impact factors, immediacy indices, and subject categories. Subject categorization remains the least objective among these indicators because the indicator is not citation-based; the ISI staff assigns journals to subjects on the basis of the journal's title, its citation patterns, etc.[1] Fields and subfields of sciences cannot (always) easily be determined. Furthermore, journals may change course over time, and journals published in different nations or with publishing houses that may not be easy to classify. In addition, journals may belong to more than one field (Boyack *et al*., 2005). The problems of classification thus raised will be especially relevant for journals that aim to function between disciplines. In contrast, we use raw citation data from the *SSCI* and *SCI*.

Analyzing the combined SSCI and SCI databases with citations among journals enables us to provide a specific characterization of the current state of heterodox economics. We do so by adopting a two-stage strategy. First, we analyze the situation for two journals—the *Journal of Economic Issues* and the *Cambridge Journal of Economics*—separately. Secondly, we combine the citation patterns extracted from the SCI and the SSCI databases for a set of journals to arrive at a more comprehensive view. At the suggestion of the guest editors of this special issue, we have included the *Cambridge Journal of Economic* (CJE), *Journal of Economic Issues* (JEI), *Journal of Post Keynesian Economics* (JPKE), *Science and Society* S&S), *Feminist Economics* (FE), and the *New Left Review* (NLR). Some links between these and other journals in the environment thus created can be expected; others may not be foreseen; and some links are stronger than expected. Other links between journals may be surprisingly absent. Links with adjacent fields may be expected for journals that aim to be interdisciplinary in particular. This does not mean that such links will materialize or have materialized.

We collected information from the databases using one journal at a time as a 'seed journal' since citation environments only make sense locally (Leydesdorff 2006a). Indeed, the citation patterns and behaviour for these journals, even though they are in roughly the same environment, can be quite different. Using methods developed in Social Network Analysis (SNA) and Scientometrics, it is possible to determine what relations in particular have actually shaped a field. We use the matrix consisting of aggregated citations among journals in this field in a number of different ways, instead of simply counting the number of citations or calculating the Impact Factor (IF) of a journal as a ranking tool.

In this study, we are primarily interested in knowledge exchanges in the field of heterodox economics, and so IFs are not as informative as measures of the extent to which a journal is centrally located in between other journals in a specific environment. Determining the extent to which a journal is indispensible in its field may thus be done by calculating its centrality. Freeman (1978/9) developed four concepts of centrality in a social network (Wasserman & Faust, 1994; Hanneman & Riddle, 2005; De Nooy *et al*., 2005). Centrality can be analyzed in terms of

1. "degrees": in- and outgoing information flows (ties, citations) from a node;
2. "closeness": the 'distance' of a node from all other agents in a network;
3. "betweenness": the extent to which a node is positioned on the shortest path between any other pair of nodes in the network; and
4. centrality in terms of the projection on the first "eigenvector" of the matrix.

For a given network, centrality measures for a node can differ depending on the measure used. *Degree* centrality is easiest to grasp as it is the number of relations a given node maintains. Degree can further be differentiated in terms of "in-degree" and "out-degree," or incoming and outgoing relations. In our citation matrix, the references in a journal measure out-degree centrality, and being cited measures in-degree. Degree centrality is often normalized as a percentage of the degrees in a network to control for scale effects.

*Betweenness* is a measure of how often a node is located on the shortest path (geodesic) between other nodes in the network. It thus measures the extent to which the node can control communication (Freeman 1978/9). Alternatively, if a node with a high level of betweenness centrality were to be deleted from a network, the network would fall apart into otherwise coherent clusters.[2] *Closeness* centrality is also defined as a proportion. First, the distance of a node from all other nodes in the network is determined. Normalization is achieved by dividing the number of other nodes by this sum total (De Nooy *et al*., 2005, p. 127). Because of normalization, closeness centrality provides a global measure about the position of a vertex in the network, while betweenness centrality is defined with reference to the local position of a vertex (node).

Principal component and factor analysis decompose a matrix in terms of the latent *eigenvectors* which determine the positions of nodes in a network. *Eigenvector* centrality uses the factor loadings on the first eigenvector as a measure. While graph analysis begins with the vectors of observable relations among nodes (Burt, 1982), factor analysis positions nodes in terms of latent dimensions of the network. For example, core-periphery relations can be made visible using graph-analytical techniques, but not by using factor-analytical ones (Wagner & Leydesdorff, 2005).

**Box: Reservations about Citations, IF, and the SSCI / SCI**

The Impact Factor (IF) for a journal in year $t$ is calculated based on a three-year period to $t$. The number of times articles published in the years $t-2$ and $t-1$ were cited in indexed journals during the year $t$ is divided by the number of "citable items" in these two preceding years. Citable items usually are articles, reviews, letters, and notes, but not editorials and obituaries (Moed, 2005). An Impact Factor is considered by many as an indicator of a journal's importance.

Obviously, one may have reservations about the kinds of data used in this paper. Not all journals are included in the database that Thomson Scientific/ISI produces. Admission is very selective (Garfield, 1990). For example, *History of Political Economy*, the *Review of Social Economy*, and the *Review of Austrian Economics* are not included (anymore) in the database. Criteria for inclusion in the database are not fully transparent (Testa, 1997). When vying for inclusion, a steep hurdle needs to be taken by a journal. A journal that vies for inclusion will need to be able to show that the journal has an impact in the field.The quasi-IF calculated as part of the procedure, however, underestimates the real IF as it does not include within-journal-self-citations that are included in the IF of journals that are already included. Citation patterns may change substantially from year to year. If a journal publishes more issues in a volume, the fluctuations tend to be smaller however.

Among these measures of centrality, the most often used is *degree* centrality. The business of science being the development and exchange of new knowledge, betweenness centrality is in several respects a more relevant indicator. It offers a measure for the extent to which a node may control the information flow within a network (cf. Freeman 1978/9; Leydesdorff 2007b) and thus is indispensable for connecting the relevant field to its wider environment. Journals that have a high impact factor may not be as indispensable for knowledge transfer as journals that have lower impact factors.

Betweenness is a relational measure. One can expect that a journal that is "between" fields or groups of journals will show a high betweenness centrality as it brokers knowledge relevant for two largely separated fields. Not necessarily belonging to a dense group, but relating them, the total citations to this journal may nevertheless be low. Closeness centrality is less dependent on relations between individual vertices between two (or more) densely connected clusters. Closeness can thus be expected to provide a 'global' measure of "multi-disciplinarity" within a set while betweenness may provide a 'local' measure of specific "inter-disciplinarity" at specific interfaces (Leydesdorff 2007b).

Centrality measures, contrary to impact factors, are sensitive to the size of both the journal and of the field. Correlations between different centrality measures can thus be spurious: a large journal (e.g., *Nature*) which one would expect to be "multidisciplinary" rather than "interdisciplinary," might generate a high betweenness centrality only because of the large number of citations to it (i.e., its high indegree centrality). Normalization of the matrix for the size of patterns of citations suppresses this effect.

There is increasing consensus in the information sciences that normalization in terms of the cosine and using the vector-space model provides the best option in the case of sparse matrices (Ahlgren *et al.*, 2003; Salton & McGill 1983; Leydesdorff & Vaughan, 2007). Since the cosine is not affected by the number of zeros in the tails, it is best used to enhance the visualization of skewed distributions. This similarity measure is defined as the cosine of the angle enclosed between two vectors *x* and *y* (Salton & McGill, 1983), as follows:

$$\cos(x,y) = \frac{\sum_{i=1}^{n} x_i y_i}{\sqrt{\sum_{i=1}^{n} x_i^2}\sqrt{\sum_{i=1}^{n} y_i^2}} = \frac{\sum_{i=1}^{n} x_i y_i}{\sqrt{(\sum_{i=1}^{n} x_i^2) * (\sum_{i=1}^{n} y_i^2)}}$$

We use a threshold level for the cosine (cosine ≥ 0.2) to present the citation patterns of locally related journals. The cosine is very similar to the Pearson correlation coefficient, except that the latter normalizes the values of the variables with reference to the arithmetic mean (Jones & Furnas, 1987), whereas the cosine normalizes with reference to the geometrical mean. Unlike the Pearson correlation coefficient, the cosine is non-metric and does not presume normality of the distribution (Ahlgren *et al*., 2003).

These measures, their further statistical elaboration, and visualization of the accompanying networks are conveniently combined in software packages like UCINet (Bonacich, 1987; Borgatti *et al*., 2002) and Pajek (De Nooy *et al*. 2005). We have used Pajek for the visualizations. The pictures prepared include citation relations among journals that contribute more than 1% to the total citations of a seed journal—"cited" and "citing," respectively. This threshold is used in order to produce readable representations.[3] The *y*-axis of the size of a node / circle indicates the logarithm of the number of cites the journal receives, while the *x*-axis corrects for within-journal "self-citations." Thus, the larger the node for a journal, the more citations it receives in this citation environment. The rounder a node, the fewer self-citations it has. Highly elliptical nodes thus can be considered as journals with a strong focus on internal knowledge development. Alternatively, it may be said that such journals have relatively little to offer to outsiders.

## 4. Two journals highlighted (JEI, CJE; 2005 and 2006)

While the journals that are cited by articles in any journal may vary from year to year, there usually is a set of core journals that reappear year after year in a journal set representing a specialty. Over a larger number of years, one may of course see gradual shifts as some journals that cite a seed journal draw closer to a journal or move out of sight. This certainly holds for journals that are well-established in a well-established field. For example, the *Cambridge Journal of Economics* (CJE) saw its 29$^{th}$ volume in 2005 and the *Journal of*

*Economic Issues* (JEI) its 39th volume. In the first stage of our analysis we focus on these two journals as single seed journals.

The *Cambridge Journal of Economics* is strongly embedded in its field. As Figure 1 indicates, it brokers knowledge between the field of economic and social geography, technology studies, and development studies. CJE's betweenness centrality measure (33%) is exceptionally high, even though it is not the largest journal in its environment in terms of citations it receives, and thus its impact factor. Surprisingly, perhaps, only two of the other journals focused on in this article—JEI and JPKE (the *Journal of Post Keynesian Economics*)— surpass the threshold imposed. CJE seems to be positioned on the edges of what may be called heterodox economics, and to take its strength from developing knowledge that is relevant for fields beyond the boundaries of the discipline of economics broadly conceived.

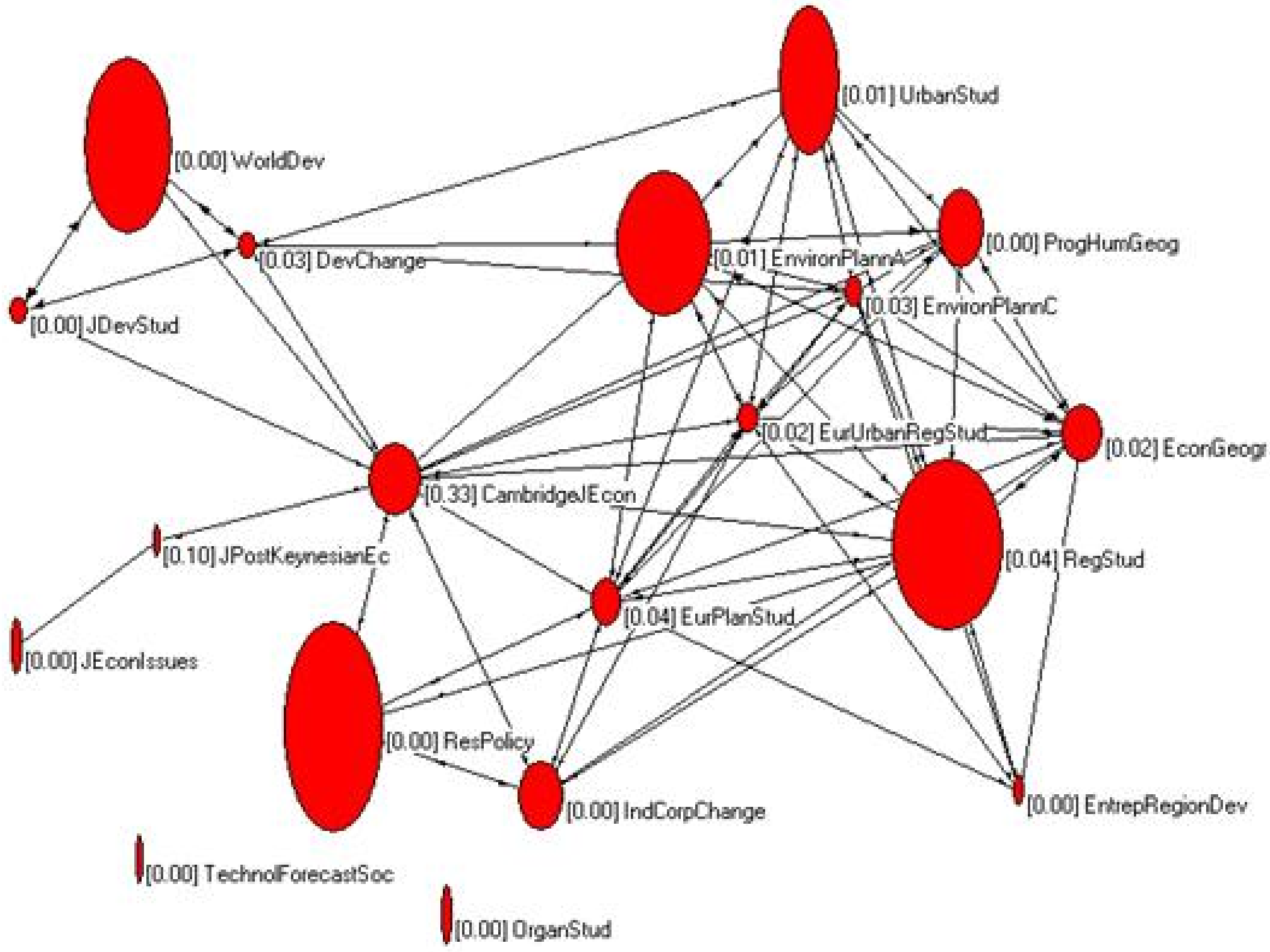

**Figure 1: Cambridge Journal of Economics**, cited, 2005.
Source: Thomson-ISI *Journal Citation Reports*, 2005; own calculations using Pajek software.

Notes: Citations between journals are normalized (cosine ≥ 0.2) to enhance visualization. Size of node represents number of citations (normalized); *x*-axis corrects for self-citations. Betweenness centrality measures for each node are included in brackets.

The citing pattern from CJE to other journals is interesting, too: of the citations to journals that surpass the threshold, only mainstream economics journals such as the *American Economic Review* (AER), the *Economic Journal*, and the *Quarterly Journal of Economics* (QJE) remain. In 2006 JPKE enters the picture too. Citations from CJE above this threshold are spread equally across these journals, both in 2005 and 2006. A large number of articles in other journals may have been cited by authors whose articles appeared in the 2005 CJE, but these citations are too dispersed to surpass the threshold, or else are to journals not included by Thomson in their lists.

The position of the JEI seems more vulnerable than that of the CJE. Its environment holds fewer journals from which it receives a substantial number of cites given this threshold (cosine ≥ 0.2), and the relations among the journals by which it is cited appear less connected, and cluster more into sub-groupings. Because of the disconnected clusters, betweenness centrality scores need to be interpreted carefully. These scores cannot be compared easily between networks.

Two larger clusters and one somewhat smaller constitute the citation impact environment of the JEI. The first cluster is that of a number of heterodox journals, including the CJE and the JPKE. The other larger cluster is made up of journals in management or organization studies. FE surpasses the threshold for inclusion in the picture as a node, but the relations it maintains are not strong enough for these to surpass the threshold for inclusion. FE thus features as a seemingly disconnected node in Figure 2.

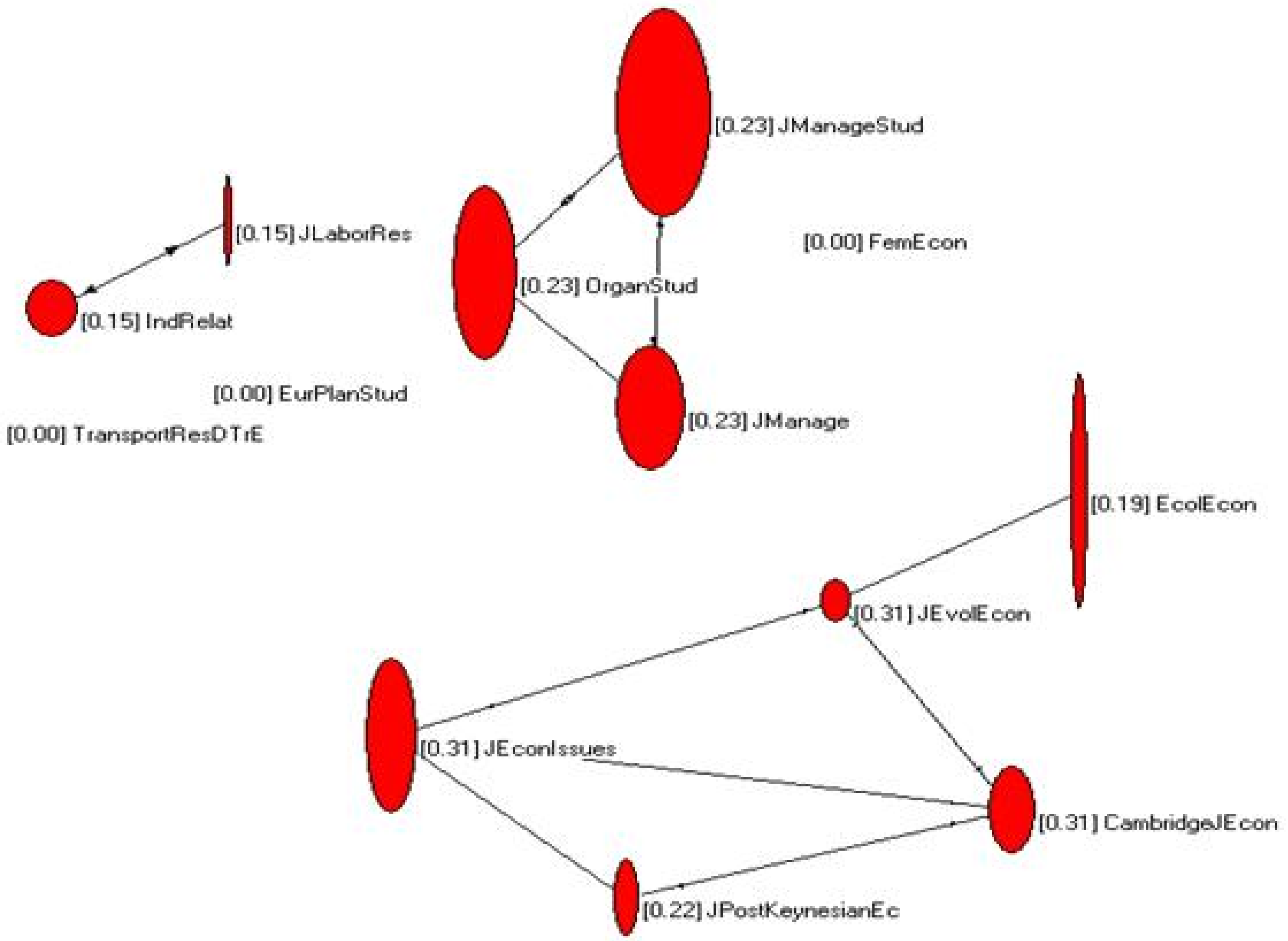

**Figure 2: Journal of Economic Issues**, cited, 2005.
Source: Thomson-ISI *Journal Citation Reports*, 2005; own calculations using Pajek software. Notes: Citations between journals are normalized (cosine ≥ 0.2) to enhance visualization. Size of node represents number of citations (normalized); *x*-axis corrects for self-citations. Betweenness centrality measures for each node are included in brackets.

The smaller cluster that focuses on labour and industrial relations may serve to make a point about the shape of the nodes. The size of the nodes indicates the numbers of citations a journal receives from other journals in the particular environment. As the *x*-axis of the node signifies the correction for self-citations, it follows that the more elliptically shaped journals cite articles in the same journals more than articles in other journals. Citation within the journal, such as for the *Journal of Labor Research*, need not be problematic since the field may be organized energetically around a journal that is developing insights at a rapid pace and so no input from relevant contexts is sought.

Citing behaviour may be very different from cited behaviour. For the JEI, only two journals received enough citations in 2005 to be included in a picture such as this: the *American Economic Review* and the *Journal of Economic Perspectives* (JEP). For 2006 the JEP is replaced by the QJE. As a sources of knowledge both the CJE and the JEI thus use mainstream economic journals as reference frames. This may be a surprising finding. Other citations from the JEI are too thinly spread or are to journals not listed in the SSCI and SCI lists.

## 5. The citation environment of heterodox econ journals

Combining the data extracted from the SSCI and SCI databases for the six selected heterodox economics journals yields the following picture.

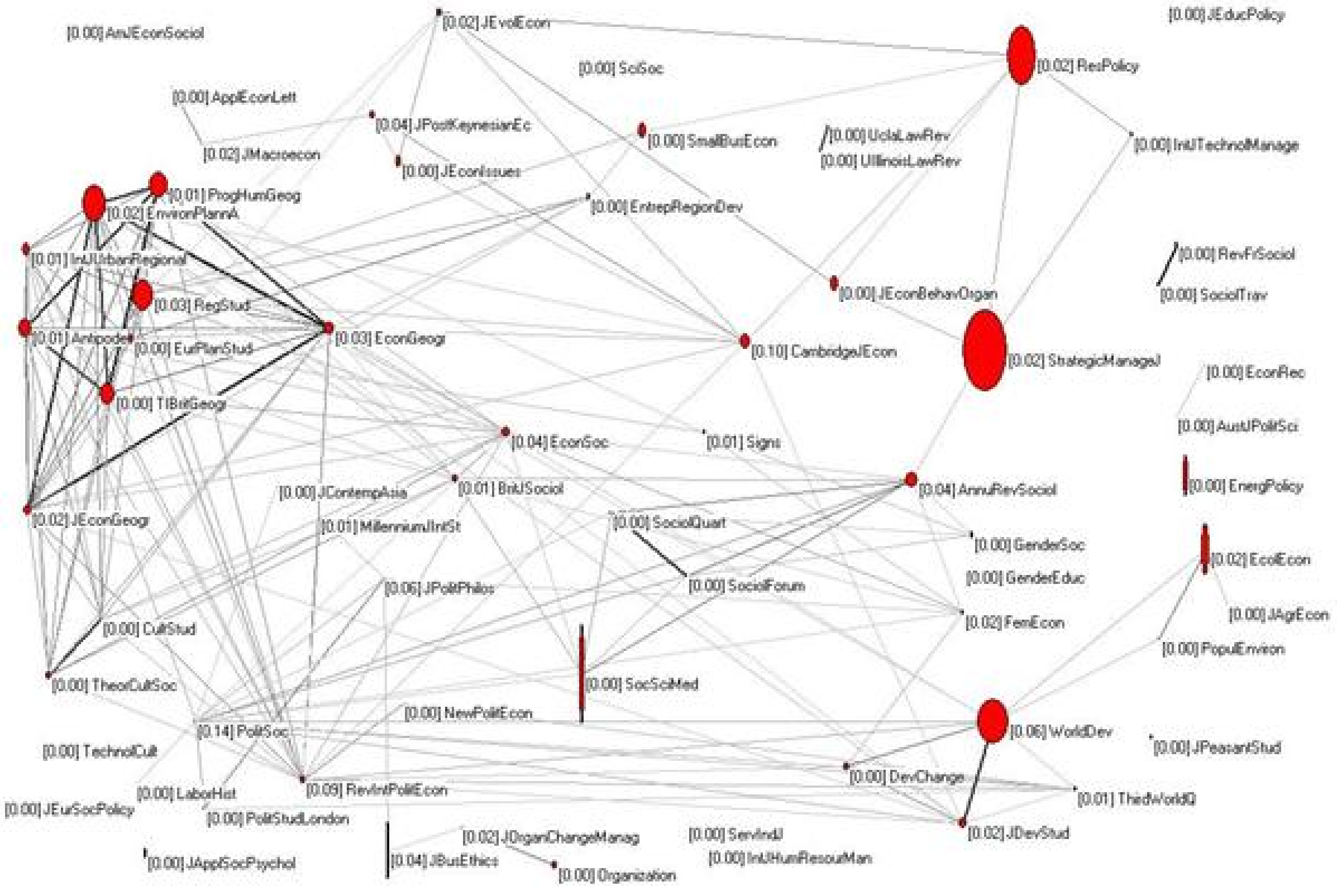

**Figure 3:** Heterodox Economics journals, cited, 2006.
Source: Thomson-ISI *Journal Citation Reports*, 2006; own calculations using Pajek software.

Notes: Citations between journals are normalized (cosine ≥ 0.2) to enhance visualization. Size of node represents number of citations (normalized); *x*-axis corrects for self-citations. Thickness of lined/edges represents tie strength.

One thing to note is that the *New Left Review* (NLR) does not even enter the picture. Despite its—for the social sciences—impressive IF of 1.115 (in 2006), none of the citations received surpasses the 1% threshold. This could reflect the editorial policy of a journal that covers the broadest possible range of themes. Citations to it would then also be spread. Analyzed as a single seed journal, it becomes apparent that it is positioned very centrally between social & human geography, political science, cultural studies and sociology in particular. This picture does not contain any of the heterodox journals included in the analysis for this article, but it does include the relatively new *Review of International Political Economy* and *New Political Economy*.

*Feminist Economics* in this representation is not classified as one of the heterodox economics journals, but it is positioned primarily in a cluster of feminist journals which are close to sociology (Woolley, 2005 and 2008; Lee, 2008). *Science & Society* does enter Figure 3, but the citations it receives from other journals never surpass the threshold maintained for the inclusion of edges so that Figure 3 remains tractable. In the terminology of social network analysis, this journal would be considered as an isolate. The picture when *Science & Society* is used as a single seed journal, however, looks much more perilous than that for NLR: it has no connections with any other journal in its own local environment to speak of.

CJE again seems relatively well positioned in Figure 3. Its betweenness centrality score is relatively high, as it relates to a diverse set of neighbouring fields. The JEI and the JPKE are situated in the periphery of this picture. Given its intellectual heritage, the JEI should be able to contribute to discussions in the fields of social & economic geography, social policy, and technology studies. These fields, but also management and organization studies, are relatively open minded and have increasingly turned to institutional theory as a source of inspiration. However, JEI has not been able so far to build bridges to these fields that are strong enough to become self-sustaining. FE, despite its relatively young age, is at present (in 2005 and 2006) better connected. FE links journals from sociology, social policy, and development studies. For 2003 and 2004—figures not included in this article, but available from the authors upon request—FE had very few connections in an at that time relatively disconnected citation environment.

**Table 1: Selected Characteristics for Selected Journals in the *Science Citation Index* and *Social Science Citation Index*, 2006**

| Journal | | Betweenness Centrality (%) | Closeness Centrality (%) | Impact Factor |
|---|---|---|---|---|
| | | (local) | | (global) |
| *Cambridge Journal of Economics* | CJE | 27 | 63 | 0.571 |
| *Feminist Economics* | FE | 13 | 43 | 0.667 |
| *Journal of Economic Issues* | JEI | 1 | 23 | 0.338 |
| *Journal of Post Keyn Economics* | JPKE | 11 | 40 | 0.244 |
| *New Left Review* | NLR | 56 | 94 | 1.115 |
| *Science & Society* | S&S | 0 | 0 | 0.364 |

Source: Thomson-ISI *Journal Citations Reports*; local centrality measures – own calculation.

A discussion of the local citation environments of journals such as those included here in the field of heterodox economics brings out clearly that the mere evaluation of journals in terms of their Impact Factors leaves important features of their academic environment untouched. Impact Factors such as those included in Table 1 for the six journals for 2006 thus offer only the barest kind of information. Yet, one should not conclude that the citation data supplied by Thomson / ISI is useless. The social network pictures that may be produced on the basis of the data, as well as measures such as betweenness centrality, offer striking insights into the position that particular journals or sets of journals find themselves in.

## 6. Some Concluding Remarks

Citations to journals, as compiled in the *Social Science Citation Index* and the *Science Citation Index*, and ranked as Impact Factors (IF) for journals, have come to be seen as increasingly important in assessing the research output of scholars despite some obvious drawbacks. Such citation information can be used to analyze journals in their local environment of relations as well, and not simply to calculate objective measures of worth. In this article we have gone much further using citation data than the mere computation of Impact Factors. Using tools developed in scientometrics and social network analysis, we are able to provide encompassing pictures and indicators of a journal's indispensibility, of a set of journals in the field of heterodox economics.

From among the six journals selected for analysis, JEI, S&S and JPKE are rather de-centrally located and their editors will have to consider attempts to sustain and strengthen the positions of these journals. Then again, a journal such as NLR has the highest Impact Factor of all, yet does not enter in the combined picture as its citation pattern is very dispersed, and perhaps purposefully and usefully so. A journal may only pursue such a strategy, it would appear, if it has a strong identity in at least one other respect. NLR's identity is its (far) left-of-center political stance. While being also left-of-center, the position that S&S takes seems much more perilous.

While the IF of CJE hovers in the range of 0.4 – 0.7, it does maintain a strong position in the field that it will not lose quickly. *Feminist Economics* has seen its IF rise rapidly over the years apparently by not aligning very closely to heterodox economics, but instead moving into the interface between (mainstream) economics, gender studies, and sociology.

## References

Ahlgren, P., Jarneving, B., & Rousseau, R. (2003) "Requirement for a Cocitation Similarity Measure, with Special Reference to Pearson's Correlation Coefficient" *Journal of the American Society for Information Science and Technology* 54(6): 550-560.

Bensman, S. J. (2007). Garfield and the Impact Factor. *Annual Review of Information Science and Technology,* 41, 93-155.

Blank, R.M. (1991) "The Effects of Double-Blind versus Single-Blind Reviewing: Experimental Evidence from *The American Economic Review*" *American Economic Review* 81(5): 1041-1067.

Bonacich, P. (1987) "Power and Centrality: A Family of Measures" *American Journal of Sociology* 92(5): 1170-1182.

Borgatti, S. P., Everett, M. G., & Freeman, L. C. (2002) *UCINet for Windows: Software for Social Network Analysis*. Harvard, Cambridge, MT: Analytic Technologies.

Bourdieu, P. (1996 [1992]) *Rules of Art: Genesis and Structure of the Literary Field*. Stanford, CA: Standford University Press.

Boyack, K. W., Klavans, R., & Börner, K. (2005). "Mapping the Backbone of Science" *Scientometrics* 64(3): 351-374.

Burt, R. S. (1982). *Toward a Structural Theory of Action*. New York, etc.: Academic Press.

Freeman, L. C. (1978/1979) Centrality in Social Networks. Conceptual Clarification. *Social Networks* 1:215-239.

Garfield, E. (1972) Citation Analysis as a Tool in Journal Evaluation. *Science* 178(4060):

471-9.

Garfield, E. (1990). How ISI selects Journals for Coverage: Quantitative and Qualitative Considerations. *Current Contents*(28 May), 5-13.

Garfield, E. (Ed.) (1980) “SCI Journal Citations Reports: A bibliometric analysis of science journals in the ISI data base” Science Citation Index 1979 annual, v. 14. Philadelphia: Institute for Scientific Information.

Gibbons, M., Limoges, C., Nowotny, H., Schwartzman, S., Scott, P., & Trow, M. (1994). *The new production of knowledge: the dynamics of science and research in contemporary societies*. London: Sage.

Goldberg, P. (1968). Are some women prejudiced against women? *Trans-Action,* 5: 28–30.

Goldstone, R., & Leydesdorff, L. (2006). The Import and Export of *Cognitive Science*. *Cognitive Science,* 30(6), 983-993.

Hirsch, J. E. (2005). An index to quantify an individual's scientific research output. *Proceedings of the National Academy of Sciences of the USA,* 102(46), 16569-16572.

Lee, F. (2008). A Comment on “The Citation Impact of Feminist Economics". *Feminist Economics,*14(1), 137.

Leydesdorff, L. (2006). Can Scientific Journals be Classified in Terms of Aggregated Journal-Journal Citation Relations using the Journal Citations Reports? *Journal of the American Society for Information Science & Technology,* 57(5), 601-613.

----, (2006b) *The Knowledge-Based Economy Modeled, Measured, Simulated.* Boca Raton: Universal Publishers.

----, (2007a) Mapping interdisciplinarity at the interfaces between the *Science Citation Index* and the *Social Science Citation Index. Scientometrics* 71(3): 391-405.

----, (2007b) “Betweenness Centrality” as an Indicator of the “Interdisciplinarity” of Scientific Journals. *Journal of the American Society for Information Science and*

*Technology*.

----, (2008). *Caveats* for the Use of Citation Indicators in Research and Journal Evaluation. *Journal of the American Society for Information Science and Technology,* 59(2), 278-287.

Moed, H. F. (2005). *Citation Analysis in Research Evaluation*. Dordrecht: Springer.

Pendlebury , D. A. (2007). Article Titled "Show me the data", *Journal of Cell Biology*, Vol. 179, No. 6, 1091 – 1092, 17 December 2007 (doi:10.1083/jcb.200711140) is Misleading and Inaccurate. http://scientific.thomson.com/citationimpactforum/8427045/ (accessed January 4, 2008).

Price, D. J. de Solla (1965). Networks of scientific papers. *Science*, 149, 510- 515.

Price, D. J. de Solla (1970). Citation Measures of Hard Science, Soft Science, Technology, and Nonscience. In C. E. Nelson & D. K. Pollock (Eds.), *Communication among Scientists and Engineers* (pp. 3-22). Lexington, MA: Heath.

Rossner, M., Epps, H. V., & Hills, E. (2007). Show me the data. *Journal of Cell Biology,* 179(6), 1091-1092.

Rossner, M., Van Epps, H., & Hill, E. (2008). Irreproducible Results—a Response to Thomson Scientific. *The Journal of General Physiology,* 131(2), 183-184.

Salton, G., & McGill, M. J. (1983) *Introduction to Modern Information Retrieval*. Auckland, etc.: McGraw-Hill.

Snodgrass, R. (2006). Single- versus Double-Blind Reviewing: An Analysis of the Literature, *SIGMOD Record* 35(3): 8-21.

Testa, J. (1997). The ISI database: The journal selection process [online]. Available at http://scientific.thomson.com/free/essays/selectionofmaterial/journalselection/

Wagner, C. S., & Leydesdorff, L. (2005). Network Structure, Self-Organization and the Growth of International Collaboration in Science, *Research Policy* 34(10), 1608-1618.

Wasserman, S., & Faust, K. (1994). *Social Network Analysis: Methods and Applications*. New York, etc.: Cambridge University Press.

Whitley, R. D. (1984). *The Intellectual and Social Organization of the Sciences*. Oxford: Oxford University Press.

Woolley, F. (2005). The citation impact of feminist economics. *Feminist Economics,* 11(3), 85-106.

Woolley, F. (2008). Reply to Frederic Lee's Comment on "The Citation Impact of Feminist Economics". *Feminist Economics,* 14(1), 143-145.

---

[1] In bibliometric research, journals can be grouped either using the ISI subject categories (e.g., Leeuwen & Tijssen, 2000; Morillo *et al.*, 2003) or on the basis of clustering citation matrices (Doreian & Farraro, 1985; Leydesdorff, 1986; Tijssen *et al.*, 1987).

[2] Betweenness centrality is normalized by definition as the proportion of all geodesics that include the vertex under study. If *gij* is the number of geodesic paths between *i* and *j,* and *gikj* is the number of these geodesics that pass through *k*, *k*'s betweenness centrality is:

$$\sum_i \sum_j \frac{g_{ikj}}{g_{ij}}, \quad i \neq j \neq k$$

[3] Extraction of a relevant set of journals in a journal's neighbourhood is given by including all journals cited or citing by the specific journal in one's initial analysis. Choice of a seed journal is formal and does not affect the measures calculated in this study.